# Length-scale dependent average structures, piezoelectricity enhancement and depolarization mechanisms in a non-MPB high-performance piezoelectric alloy system $PbTiO_3$-$Bi(Zr_{1/2}Ni_{1/2})O_3$

Rishikesh Pandey[1], Dipak Kumar Khatua[1], Shekhar Tyagi[2], Mulualem Abebe[1], Bastola Narayan[1], Vasant Sathe[2] and Rajeev Ranjan[1*]

[1]Department of Materials Engineering, Indian Institute of Science, Bangalore 560012, India

[2]UGC-DAE Consortium for Scientific Research, University Campus, Khandwa Road, Indore-452001, India

## Abstract

There is a general perception that large piezoelectric response in ferroelectric alloys requires tuning the system towards a morphotropic phase boundary (MPB), i.e., a composition driven inter-ferroelectric instability. This correlation has received theoretical support from models which emphasize on field driven polarization rotation and/or inter-ferroelectric transformations. Here we show that high piezoelectric response (comparable to conventional MPB systems), both in the weak and strong field regime, can be realized even in non-MPB alloy systems. We demonstrate this in a low-Pb piezoelectric alloy system (1-x)$PbTiO_3$-(x)$Bi(Zr_{1/2}Ni_{1/2})O_3$ (PT-BNZ) exhibiting large $d_{33}$ ~ 400 pC/N and electrostrain ~ 0.5 %. We carried out a comprehensive study involving electric-field and temperature dependent X-ray powder diffraction, Raman spectroscopy, dielectric, piezoelectric and high field electrostrain measurements. Detailed analysis revealed that the composition exhibiting large piezoelectric response in our system (x=0.41) comprise of tetragonal regions of long and short-range coherence, which on the global scale gives the impression of coexistence of tetragonal and cubic-like phases. Our system is therefore qualitatively different from the conventional MPB systems such as $Pb(Zr,Ti)O_3$, $Pb(Mg_{1/3}Nb_{2/3})O_3$-$PbTiO_3$, etc., which exhibits coexisting tetragonal + rhombohedral/monoclinic phases in thermodynamic equilibrium. We found that poling-field irreversibly suppresses the cubic-like phase at room temperature, which reappears when the system depolarizes thermally due to onset of a lattice instability well before the dielectric maximum temperature. In the absence of inter-ferroelectric instability as a phenomenon, field induced polarization-rotation and inter-ferroelectric transformation are no longer plausible mechanisms to explain the large piezoelectric response in our system. The large piezoelectricity in our case is solely due to enhanced mobility of the tetragonal domain walls enabled by domain miniaturization. We compare and discuss our results with conventional MPB systems, both in the normal ferroelectric and relaxor-ferroelectric categories. Our study proves that attainment of large piezoelectricity does not require inter-ferroelectric instability as a necessary criterion, and that the alloy category represented by the general formula $PbTiO_3$-$Bi(M'M'')O_3$ can be ideal systems for design of high performance non-MPB based piezoceramics.

**E-mail: rajeev@iisc.ac.in*

## I. INTRODUCTION

Ever since its discovery five decades ago, Pb(Zr, Ti)$O_3$ (PZT) based alloys have been the most sought after sensing and actuator materials for use in wide ranging applications in the field of space, military, automotive industry, ultrasound based medical imaging, etc. Increased environmental concerns in the last decade and half have stimulated search for Pb-free alternatives. While this push has led to the discovery of new lead-free alloys exhibiting large piezoelectric response in $BaTiO_3$-based [1] and $K_{0.5}Na_{0.5}NbO_3$-based [2,3] systems, factors such as low Curie point (as in the case of $BaTiO_3$-based compositions), and difficulty in the reproducing the desired phases (and hence properties) due to extreme sensitivity to slight changes in synthesis conditions (as in the case of (K,Na)$NbO_3$-based systems), are important hurdles regarding the commercial acceptability of these new lead-free materials. Also, the non-vertical nature of the morphotropic/polymorphic phase boundary (MPB) of these systems imparts great deal of temperature sensitivity to the piezoelectric properties, which is not desirable from a device point of view. In this scenario, it is worthwhile to explore other alternative systems, including materials with reduced Pb content. Here, by low-lead piezoelectric alloys we mean that the A-site of the perovskite structure has significant fraction of non-Pb cations, which can be contrasted with the all-lead piezoelectric systems such as $Pb(Zr_xTi_{1-x})O_3$ (PZT), $Pb(Mg_{1/3}Nb_{2/3})_{1-x}Ti_xO_3$ (PMN-PT) and $Pb(Zn_{1/3}Nb_{2/3})_{1-x}Ti_xO_3$ (PZN-PT). Alloys with the general formula $PbTiO_3$-Bi(M′M″)$O_3$ (PT-BM′M″) can be attractive candidates in this regard, as they usually exhibit high Curie point, can be compositionally tailored to exhibit very good piezoelectric response, and are easy to make with reproducible properties [4-7].

Unlike the all-Pb piezoelectric alloys where in the end members such as $PbTiO_3$, $PbZrO_3$, $Pb(Mg_{1/3}Nb_{2/3})O_3$, $Pb(Zn_{1/3}Nb_{2/3})O_3$, can separately crystallize as perovskites phase when synthesized under ambient pressure conditions, the component Bi(M′M″)$O_3$ in PT-BM′M″ do not crystallize in perovskite structure under ambient pressure synthesis conditions. The perovskite phase is rather formed only when they are synthesized at high pressure (~ 5 - 6 GPa) and high temperatures (~ 1000 $^oC$) [8-10]. However, for the sake of convenience, we may treat Bi(M′M″)$O_3$ as a virtual compound which can be alloyed with other real perovskite compounds. The solubility of several BM′M″ is considerably high in $PbTiO_3$. Consequently, it can induce interesting changes in the crystal structure, ferroelectric and piezoelectric properties of $PbTiO_3$ [11-20]. Examples include $Bi(Ni_{1/2}Ti_{1/2})O_3$–$PbTiO_3$ (BNT-PT) [11], $Bi(Mg_{1/2}Ti_{1/2})O_3$–$PbTiO_3$ (BMT-PT) [12,17,19], $Bi(Mg_{1/2}Zr_{1/2})O_3$–$PbTiO_3$ (BMZ-PT) [13,17,19], $Bi(Zn_{1/2}Ti_{1/2})O_3$–$PbTiO_3$ (BZT-PT) [14,18,19], $Bi(Zn_{1/2}Zr_{1/2})O_3$–$PbTiO_3$ (BZZ-PT) [14,19], $Bi(Zn_{1/2}Sn_{1/2})O_3$–$PbTiO_3$ [14], $Bi(Zn_{3/4}W_{1/4})O_3$–$PbTiO_3$ [15,18], $Bi(Mg_{3/4}W_{1/4})O_3$–$PbTiO_3$ (BMW-PT) [16], $Bi(Ni_{2/3}Nb_{1/3})O_3$–$PbTiO_3$ [20] etc. From the structural standpoint, the different PT-BM′M″ alloy systems listed above can be categorized in three groups, depending on how they influence the tetragonality (c/a = 1.06) of $PbTiO_3$. While $Bi(Zn_{1/2}Ti_{1/2})O_3$ and $Bi(Zn_{3/4}W_{1/4})O_3$ increases the tetragonality of $PbTiO_3$ [14,15], others such as $Bi(Zn_{1/2}Zr_{1/2})O_3$ and $Bi(Zn_{1/2}Sn_{1/2})O_3$ do not affect the tetragonality [14]. Majority BM′M″, however, decreases the tetragonality with increasing alloying content. First principles calculations by Grinberg and Rappe [19] have shown that the overall polarization and tetragonality of PT-BM′M″ depends on the coupling between the off-centred displacement of ions on the A-site (comprising of Pb and Bi ions) and those on the B-site. While Bi substituting the Pb-site in $PbTiO_3$ invariably enhances the ferroelectric distortion, the overall polarization and the tetragonal distortion of the alloy is determined by the type of elements M′ and M″ replacing Ti on the B-site [19].

For PT-BM′M″ alloys to be exhibiting good piezoelectric response, it is important that not only the tetragonality decreases with increasing $Bi(M'M'')O_3$ concentration, but also the solid solubility range should be sufficient enough to actualize the MPB like scenario. Among the Bi/Pb-based solid solution $PbTiO_3$-$BiScO_3$ (PT-BS) ($d_{33}$ ~ 450 pC/N) [21,22] and $PbTiO_3$-$Bi(Ni_{1/2}Hf_{1/2})O_3$ ($d_{33}$ ~ 446 pC/N) [23] have shown high $d_{33}$. However, both $Sc_2O_3$ and $HfO_2$ required for the synthesis of these alloys are costly, making these alloy systems unattractive for mass production and commercial application. Recently, a new alloy system in this category namely (1-x)$PbTiO_3$-(x)$Bi(Ni_{1/2}Zr_{1/2})O_3$ ((1-x)PT-(x)BNZ) is reported to show high $d_{33}$ ~ 400 pC/N [6], making it a very attractive piezoelectric material in the low-Pb category. A perusal of literature suggests an important difference between the MPB of $PbTiO_3$-$Bi(M'M'')O_3$ and those of the conventional MPB piezoelectrics. The MPB compositions of the conventional systems such as PZT and PMN-PT exhibit coexistence of ferroelectric phases with tetragonal and rhombohedral/monoclinic symmetries [24,25]. In contrast, the coexisting phases in $PbTiO_3$-$Bi(M'M'')O_3$ are most often reported as tetragonal and "cubic-like" [11-13,17]. The term "cubic-like" structure has been used to highlight the fact that the actual structure on the local scale may not be truly cubic, but it appears to be so on the global scale (as in X-ray diffraction based structural studies). Based on the analogy with the conventional MPB systems, some groups have assigned rhombohedral ($R3m$) symmetry to the cubic-like phase [6,11-13]. Others have assigned cubic structure [17].

The understanding of the nature of the cubic-like phase is of great significance, as it has great implication on our understanding of the mechanism(s) associated with the large piezoelectric response in this system. The two prominent mechanisms often invoked to explain the high piezoelectric of the conventional MPB systems are: (i) polarization rotation, which relates the enhanced piezoelectric response to ease of polarization rotation inside the unit cell on application of electric field [26-30], and (ii) enhanced domain wall mobility [31-34]. Some reports in the recent past have attributed field driven structural transformations as the dominant contributing mechanism [22,35-37]. The polarization rotation and the field driven inter-ferroelectric transformation mechanisms rely on the alloy's proneness to exhibit composition/temperature induced inter-ferroelectric instability, which manifests as coexistence of two ferroelectric phases in the MPB region. Electric field can change the relative fraction of the coexisting phases, which contributes to the electrostrain response [35,36]. If, as proposed by some groups, the coexisting cubic-like phase in our system has rhombohedral/monoclinic symmetry, then we can anticipate the electric field induced change in their relative fraction and, in analogy with PZT, the large piezoelectric and electrostrain in our system can be associated to the field induced inter-ferroelectric transformation. If, on the other hand, the cubic-like phase is not rhombohedal/monoclinic, then we cannot invoke these mechanisms as plausible. The large piezoelectric of our model system (1-x)$PbTiO_3$-(x)$Bi(Ni_{1/2}Zr_{1/2})O_3$ ($d_{33}$ ~ 400 *pC/N*) [6] provides an opportunity to examine if polarization rotation and/or field induced structural transformations are necessary mechanisms to achieve large piezoelectric and electrostrain response in piezoelectric alloys or not. We carried out a comprehensive investigation using complementary experimental techniques involving electric-field and temperature dependent X-ray diffraction (XRD), Raman spectroscopy, dielectric and piezoelectric measurements. We found that the entire composition region exhibiting two-phase state (cubic-like + tetragonal) before poling transforms to tetragonal after poling. Interestingly, the Raman spectra, which probes structural coherence on the local scale, do not show a corresponding dramatic change after poling, thereby suggesting that on the local scale, the structure is still tetragonal in the cubic-like phase. The poling induced cubic to tetragonal transformation is therefore a manifestation of the increase in the coherence length of the tetragonal regions, making XRD reveal the true symmetry (tetragonal) present on the local scale. Our study reveals that exhibiting significant enhancement in the piezoelectric response occurs in our system without the existence of polarization rotation/phase transformation mechanisms. The large piezoelectric and electrostrain response of our system is primarily due to enhanced motion of tetragonal domain walls.

## II. EXPERIMENTAL DETAILS

(1-x)PT-xBNZ solid solutions were synthesized via conventional solid-state ceramic route. High-purity analytical-reagents (AR) grade $Bi_2O_3$ (99 %), PbO (99.9 %) , NiO (99 %), $ZrO_2$ (99.99 %) and $TiO_2$ (99.8 %) chemicals from Alfa Aesar were wet milled according to stoichiometric proportions in a agate jar with agate balls and acetone as the mixing media for 6 h using a planetary ball mill (Fritsch P5). The thoroughly mixed powder was calcined at 850 °C for 6 h. The calcined powder was mixed with 2 wt% polyvinyl alcohol-water solution and pressed into form of disks of 15 mm diameter by using uniaxial die at 8 ton. Sintering of the pellets was carried out between 1100 °C - 1150 °C for 3 h in closed alumina crucible. Calcined powder of same composition was kept inside the crucible as sacrificial powder during sintering. For X-ray diffraction measurements, sintered pellets were crushed into fine powder and annealed at 500 $^{o}$C for 6 h to remove the strains introduced during crushing. X-ray powder diffraction (XRD) data was collected using a Rigaku (SMART LAB, Japan) diffractometer with a Johanson monochromator in the incident beam to remove the Cu-$K\alpha_2$ radiation. Dielectric measurement was carried out using a Novocontrol (Alpha-AN, USA) impedance analyzer. For the study of electric field induced phase transformations, the sintered pellets were electroded by coating with silver paste and cured at 100 $^{o}$C for 1 h. The electroded pellets were poled at room temperature in silicone oil for 1 h by applying a DC-electric field of 30 kV/cm. Longitudinal piezoelectric coefficient ($d_{33}$) was measured by poling the pellets at room temperature for 1 h at a DC-electric field of ~30 kV/cm using piezotest PM-300. The strain loop and the polarization electric-field (*P*-*E*) hysteresis loop were measured with a Precision premier II loop tracer. High temperature Raman data was collected in the backscattering geometry using a diode laser excitation source (473 nm) coupled to a Labram-HR800 micro-Raman spectrometer equipped with a 50× objective with an appropriate edge filter and a Peltier-cooled charge coupled device detector. Linkam, UK make THMS-600 stage was used for temperature variation. Structure refinement was carried out by FULLPROF software [38].

## III. RESULTS

### A. The critical composition range

The morphotropic phase boundary (MPB) composition range of (1-x)PT-xBNZ was determined by visual inspection of the characteristic pseudo-cubic Bragg profile $\{200\}_{pc}$. This profile shows only two Bragg peaks in pure tetragonal (T) compositions (x ≤ 0.38), Fig. 1.

The MPB compositions (x = 0.39 - 0.42), on the other hand, show an additional peak in between the two tetragonal peaks. For x ≥ 0.43, the tetragonal peaks have become almost invisible, and all the peaks in the pattern appear as singlet, suggesting a cubic-like (C) structure. Whole pattern fitting of the XRD data was carried out using Rietveld method with tetragonal (*P4mm)* + cubic (*Pm*-3*m*) average structures. As evident from Fig. 1, this structural model fits the data very well. The refined structural parameters for (1-x)PT-xBNZ are listed in the Table.1. The isotropic thermal parameters ($B_{iso.}$) of the A-site cations were very large ( ~ 3.0 Å$^2$) both in the tetragonal and the cubic-like phases. We refined the anisotropic thermal parameters (U) of the A-site cations in tetragonal phase. For other atoms only the $B_{iso}$ were refined. This resulted in a overall good fit between experimental and calculated pattern. Very large thermal parameter of the A-site cations has also been reported in Pb-based relaxor systems such as $Pb(Mg_{1/3}Nb_{2/3})O_3$ (PMN) [39-43] due to local positional disorder which is a very common feature of relaxor ferroelectrics. Like our case, very significant improvement in Rietveld fits have also been reported by invoking the anisotropic thermal parameters in these systems as well [39-43].

We may note that although the compositions x = 0.44 and x= 0.38 appears as pure cubic-like and tetragonal, respectively, the fit was noticeably improved when the second minor phases was included in the structural model. Fig. 2(a) shows the composition dependence of the lattice parameters and tetragonal phase fraction (Fig. 2(b)) of (1-x)PT-xBNZ. The tetragonal a-parameter increases, and the c-parameter decreases with increasing x. The rate of change of the parameters can be seen to decrease noticeably for x > 0.40. This is more clearly revealed in the composition variation of the tetragonality (c/a) shown in Fig. 2b. Interestingly, the tetragonal phase fraction also follows the same trend, Fig. 2(b), suggesting a correlation between the two parameters.

**B. Piezoelectric and high-field electrostrain response**

Fig. 2(c) shows compositional variation of the longitudinal piezoelectric coefficient ($d_{33}$) and the MPB region (marked by the vertical dotted lines). All compositions in the MPB region show high $d_{33}$. The highest $d_{33}$ of 385 pC/N was obtained for the composition x = 0.41. Our value is close to ~ 400 pC/N, reported before for this alloy system [6]. Even x = 0.44, the composition exhibiting pure cubic-like phase (Fig. 1) shows $d_{33}$ as high as 260 pC/N, confirming that its structure must be non-cubic ferroelectric on the local length scale. This aspect becomes clear in the XRD patterns of the poled sample of x = 0.44, discussed in the next section. Fig. 3(a) shows the variation of unipolar electrostrain with electric field up to an applied field of 60 kV/cm. The composition x = 0.41 shows the maximum electrostrain of 0.42 %. This corresponds to a large signal converse piezoelectric response ($d^*_{33}$) defined as $S_{max}/E_{max}$, ~ 700 pm/V, Fig. 3(c). Our electrostrain at 60 kV/cm is even higher than what has been reported for the MPB compositions of PZT [44]. We also noted that this composition exhibits the smallest coercive field, Fig. 3(b). We succeeded in achieving electrostrain to ~ 0.5 % on x = 0.41 when the electric field amplitude was increased to 85 kV/cm, inset Fig. 3(a). Such high value of strain has not been reported on this system in earlier works [6]. Given the fact that the electrostrain is sensitive to grain size, i.e. for any given composition, the electrostrain can be reduced by reducing the grain size of the specimen, we ensured that average grain size for all the compositions were nearly the same (4 - 5 microns), Fig. 4. This confirms that the trend in the properties shown in Figs. 2 and 3 are intrinsic, i.e., due to composition variation.

**C. Poling driven irreversible structural changes on the global scale**

Fig. 5(a) shows the poling driven changes in the XRD patterns of (1-x)PT-xBNZ. Poling of the dense ceramic pellets was done at 30 kV/cm at room temperature for ~ 30 minutes. The poled pellets were then crushed into fine powders and the XRD pattern was recorded. This strategy allows us to get preferred orientation free diffraction pattern of the poled specimen, while retaining the irreversible structural changes brought about by the poling field [45-47]. The pseudocubic X-ray powder diffraction Bragg profiles $\{111\}_{pc}$ and $\{200\}_{pc}$ of the different compositions are shown in Fig. 5(a). Evidently, all the poled compositions exhibit tetragonal structure, as the cubic-like peaks (observed in the unpoled specimens, Fig. 1,) have become invisible after poling. This scenario is true even for the composition showing pure cubic-like phase (x = 0.44), Fig. 1. Refined structural parameters of poled x = 0.41 using tetragonal space group *P*4*mm* are listed in Table.1. Fig. 5(b) shows the composition dependence of the tetragonality (c/a) of poled samples. For the sake of direct comparison, we also show the tetragonality of the tetragonal phase in unpoled specimens (shown by asterisks) of the same compositions. It is interesting to note that the tetragonality of the poled specimen is considerably larger than that in the unpoled specimen in the composition range x~ 0.40 – 0.42. The difference is considerably reduced for x = 0.39, the composition exhibiting larger fraction of the tetragonal phase within the MPB region, Fig. 2(b). This further confirms a correlation between the measured tetragonality and the fraction of the tetragonal phase as pointed out in Section A.

**D. Confirmation of relaxor ferroelectricity**

Fig. 6 shows the temperature variation of the real (ε′) and imaginary (ε″) part of permittivity of x = 0.41. The broadness of the permittivity peak, shifting of the permittivity maximum to higher temperature on increasing frequency, confirm relaxor ferroelectric behaviour. Vogel-Fulcher analysis (inset Fig. 6) of the frequency dependence of the imaginary part of permittivity (ε″) maximum temperature [48] suggests activation energy ($E_a$), relaxation time ($\tau_o$) and the Vogel-Fulcher freezing temperature ($T_f$) as 3.38 x $10^{-3}$ eV, 2.08 x $10^{-7}$ sec., and ~ 257 $^oC$, respectively. Figs. 7(a) and (b) compare the temperature variation of the dielectric permittivity of unpoled and poled specimens, respectively of PT-BNZ (x = 0.41). A notable distinction is the occurrence of a small permittivity peak at ~ 120 $^oC$ in the poled sample (Fig.7(b)), which is not visible in the unpoled specimen, Fig. 7(a). We assign this temperature as the depolarization temperature ($T_d$) since the $d_{33}$ signal became almost zero when the poled pellet was annealed at this temperature, Fig.7(c). A significant departure from the Curie-Weiss behaviour was also found below 470 $^oC$ (Fig. 7(d)), suggesting this to be the Burn's temperature of this system below which polar nano regions are expected to appear [49].

**E. Structural change on the global scale during depolarization**

To understand the nature of structural changes associated with the thermal depolarization on the global scale, we carried out high temperature XRD measurement on poled PT-BNZ (x = 0.41). For this, the poled pellet was ground to powder to get rid of preferred orientation effect in the powder diffraction pattern. The structure of the poled specimen is tetragonal at room temperature, Fig. 8(a). On heating, separation between the two $\{200\}_{pc}$ peaks decreases continuously. Onset of a new peak corresponding to the cubic-like phase can be seen in between the two tetragonal peaks at 130 $^oC$. The cubic-like and the tetragonal phases coexists up to 150 $^oC$. The structure appears completely cubic at 200 $^oC$, Fig. 8(a). This proves that the sharp drop in the $d_{33}$ above 100 $^oC$ (Fig. 7(c)) is associated with the appearance of the cubic-like phase during heating of the poled sample. Figs. 8(b), (c) and (d) show the temperature variation of the lattice parameters, unit-cell volume and tetragonality. We find weak anomaly in the temperature dependence of cell volume near the depolarization temperature ~ 150 $^oC$.

**F. Local structure and lattice instability: Raman study**

Fig. 9(a) shows a Lorentzian fitted Raman spectrum of the x = 0.41. The Raman modes were assigned following earlier studies [50,51]. Due to instrumental limitations we did not consider peaks below 100 $cm^{-1}$ for analysis. The first peak in our spectrum appears at ~ 180 $cm^{-1}$, which is assigned as the $A_1$(1TO) mode. The other peaks are assigned as E(2TO) ~ 227 $cm^{-1}$, ($B_1$+E) ~ 277 $cm^{-1}$, $A_1$(2TO) ~ 320 $cm^{-1}$, E(3TO) ~ 488 $cm^{-1}$, $A_1$(3TO) ~ 585 $cm^{-1}$, E(3LO) ~ 700 $cm^{-1}$ and $A_1$(3LO) ~ 760 $cm^{-1}$. With respect to the parent compound $PbTiO_3$, the $A_1$(1TO) is considerably hardened in our system (it increases by ~ 28 $cm^{-1}$, i.e. from ~ 152 $cm^{-1}$ in $PbTiO_3$ to ~ 180 $cm^{-1}$). A perusal of literature suggests that this to be a common feature of most Bi-substituted Pb-based perovskites. For example, the $A_1$(1TO) mode in is reported at ~ 185 $cm^{-1}$ for $0.4PbTiO_3$-$0.6BiFeO_3$ [52], at ~ 172 $cm^{-1}$ for $0.65PbTiO_3$-$0.35Bi(Zn_{1/2}Ti_{1/2})O_3$ [52], and ~ 185 $cm^{-1}$ in $0.66PbTiO_3$-$0.34BiScO_3$ [53]. The $A_1$(1TO) mode is associated with off centre polar displacement of A-site cations ($Bi^{3+}$/$Pb^{2+}$) with respect to $BO_6$ octahedra in A-$BO_3$ translational mode vibration [51].

Fig. 9(b) compares the Raman spectra of the poled and unpoled specimens of x = 0.41 at room temperature. In contrast to the remarkable changes in the XRD pattern after poling (Fig. 5(a)), there are no such remarkable changes in the Raman spectra of the poled sample. Since Raman probes structural coherence on the local length scale, the nearly identical spectra of the cubic-like (before poling) and tetragonal (after poling) phases suggests that the local structure of the tetragonal and the cubic-like phases are similar. On careful examination, we however noted a slight decrease in the intensity of the mode at ~ 320 $cm^{-1}$ in the poled sample. We measured the Raman spectra of the poled specimen on first heating up to 400 $^oC$, Fig. 10(a) and then while cooling. Since the specimen was heated well above the dielectric maximum temperature (~ 270 $^oC$, Fig. 6(a)), the Raman spectra recorded during the cooling cycle can be regarded as that of the unpoled specimen. Fig. 10(b) shows the intensity of the mode at ~ 320 $cm^{-1}$ during the heating and cooling runs. The poling induced loss in intensity of this mode (Fig. 10(b)), recovered during the cooling cycle below 150 $^oC$ (Fig. 10(d)), i.e., below the depolarization temperature. Another notable feature is the softening of $A_1$(1TO) Raman mode ~ 180 $cm^{-1}$ (Fig. 10(c)) on heating, and it’s becoming invisible above 200 $^oC$ (Fig. 10(a)), confirming that the depolarization is associated with a lattice instability.

## IV. DISCUSSION

### *Conventional-MPB versus pseudo-MPB*

The large enhancement in the piezoelectric response of PT-BNZ for compositions exhibiting coexistence of tetragonal and cubic-like phases is analogous to the conventional MPB systems exhibiting large piezoelectric response for compositions showing coexistence of tetragonal + rhombohedral/monoclinic phases. From this analogy, one may be tempted to argue that the cubic-like phase could be rhombohedral/monoclinic. Such a proposition has indeed been made by some research groups in the past [6]. This view may also be emboldened by the fact that a related system $PbTiO_3$-$BiScO_3$ (PT-BS) shows clear evidence of rhombohedral/monoclinic structure (the splitting of pseudo-cubic $\{111\}_{pc}$ Bragg profile into two) outside the MPB region on the BS excess side [5,22]. Thus, the nature of MPB in PT-BS is like the MPB in PZT [24], i.e. representing a boundary separating tetragonal and rhombohedral phase fields in the composition-temperature phase diagram. To get better insight regarding the nature of the cubic-like phase, it is worthwhile to examine how poling affects the structural states in conventional MPB systems characterized by tetragonal + rhombohedral/monoclinic, and the non-conventional ones such as ours. In the case of PZT, poling of the MPB composition decreases the tetragonal phase by ~15 - 20 % [35]. A similar change has been reported in the MPB composition of PT-BS [22], with the difference that rhombohedral/monoclinic (instead of tetragonal) phase is suppressed by ~20 %. These observations, suggest that poling merely alters the relative fraction of the coexisting phases in conventional MPB systems, and does not eliminate one of the coexisting phases altogether as we see in our case (Fig. 5). Another notable difference is that the fraction of transformed phase decreases as the system approaches the rhombohedral/monoclinic boundary in conventional MPB systems. For example, the rhombohedral composition just outside the MPB of the PT-BS system does not show poling induced change in structure [5]. The equivalent composition in our alloy system ($x = 0.44$), on the other hand, transforms to tetragonal after poling (Fig. 5). The inability of the poling field to eliminate one of the coexisting phases altogether either in PZT or PT-BS confirms that the coexisting phases are structurally genuine, and in thermodynamic equilibrium. The characteristic Raman modes of rhombohedral/monoclinic and tetragonal symmetries have been identified in Raman spectrum of the MPB composition of PZT [54], lending authenticity to their existence as a genuine structural phase. The complete elimination of the coexisting cubic-like phase in our system suggests that it does not represent a genuine structural state. As our Raman study suggests, the cubic-like phase is merely a manifestation of coherent scattering from small tetragonal domains on the global scale. The non-conventional MPB therefore represents coexistence of short and long range tetragonal domains. This scenario get support from the results of local structure analysis using pair distribution function for the cubic-like phase of the alloys system $BaTiO_3$– $Bi(Zn_{1/2}Ti_{1/2})O_3$ [55]. Usher et al [55] have shown that the average structure appears tetragonal on smaller length scale and cubic on larger length scale. The decrease in the measured tetragonality with decreasing fraction of the tetragonal phase (Fig. 2b) in the unpoled specimens, and increase of tetragonality after poling (i.e., after supressing the cubic-like phase, Fig. 5b) is consistent with the observation of Usher et al [55]. In this scenario, the increase in the coherence length of the tetragonal domains after poling will make the tetragonality appear to increase on the global scale, as we have seen (Fig. 5(a)).

***Piezoelectricity enhancement mechanism in the pseudo-MPB***

Our conclusion that the cubic-like phase is not rhombohedral/monoclinic but tetragonal, has important implications. Polarization rotation [27] and the field induced inter-ferroelectric transformations [35,36], which have often been invoked to explain the high enhanced piezoelectric response in conventional MPB systems (e.g. PZT), is no longer applicable for our non-conventional MPB alloy. Since the MPB composition of our system is characterized by existence of short and long range tetragonal domains in the unpoled state, the significant enhancement of the piezoelectric response and the electrostrain can only be associated with the ease of domain wall displacement [31-34]. Very recently, Abebe et al have demonstrated that composition showing large piezoelectric response (> 500 pC/N) in the $(Ba,Ca)(Ti,Sn)O_3$ is the one which exhibit a combination of good polarization and easy motion of domain walls [56]. This view is also consistent with our results since the coercive field is minimum for composition exhibiting highest piezoelectric response in our system. There is therefore a correlation between ease of domain wall motion and onset of the cubic-like phase, i.e., when the system starts exhibiting tetragonal ferroelectric domains of very short coherence length. In such a scenario, a large electrostrain is possible if the system has a combination of large tetragonality and high domain mobility. However, it is generally observed that domain switching becomes increasingly difficult with increasing tetragonality of the ferroelectric phase. In view of this the composition exhibiting highest electrostrain would be the one showing optimum combination of tetragonality and domain mobility.

Another important factor in this regard is the extent of the reverse switching when the field is switched off [57,58].

***Comparison with the cubic-like phase in $Na_{1/2}Bi_{1/2}TiO_3$-$BaTiO_3$***

It is worth comparing origin of cubic-like phase in our system PT-BNZ and that in the well-known lead-free piezoelectric $0.94Na_{1/2}Bi_{1/2}TiO_3$-$0.06BaTiO_3$ (NBT-6BT) [47,59-60]. The cubic-like phase of NBT-6BT evolves from the monoclinic average structure of $Na_{1/2}Bi_{1/2}TiO_3$ (NBT) [61,62]. On poling it transforms to rhombohedral [60] or tetragonal + rhombohedral [47]. In contrast to $PbTiO_3$, which is well behaved the classical ferroelectric, NBT exhibits high degree of positional disorder on the A site (randomly occupied by Na and Bi cations), and local in-phase octahedral tilt [63-68] which is incompatible with a long-range ferroelectric order. The intrinsic positional disorder in NBT arises due to the qualitatively very different characteristics of the Na-O and Bi-O bonds [66]. Thus, in contrast to $PbTiO_3$, the monoclinic (*Cc*) average structure of NBT is not a thermodynamically stable structure. Rao et al have shown that poling supresses the positional disorders, making the thermodynamically stable rhombohedral structure (space group *R*3*c*) reveal itself on the global scale [45,46,67]. The onset of the cubic-like phase in the unpoled state of NBT-6BT is a consequence of adding further disorder in the already (intrinsically) positionally disordered parent compound, NBT. This makes NBT-6BT a strong relaxor ferroelectric with significantly enhanced dielectric dispersion suggesting great deal of structural heterogeneity [68]. However, it is important to note that although both NBT-6BT and our system exhibit cubic-like phase on the global scale, the piezoelectric response of the NBT-6BT is significantly low (~ 190 pC/N) as compared to ours. Eerd et al [69] have argued that the low $d_{33}$ of NBT-6BT is because the system retains its local structural correlations even above the depolarization temperature, i.e., the depolarization in NBT-6BT is not thermodynamic instability. However, our system does exhibit large piezoelectric ($d_{33}$ ~ 400 pC/N) and electrostrain response (~ 0.5 % at 85 kV/cm) even though it also retains its local structure above the depolarization temperature (the Raman spectra of PT-BNZ do not change dramatically above the depolarization temperature (Fig. 10(a)). This implies that triple point may be helpful, but need not be fundamental necessity to achieve large piezoelectric response. In any case, the arguments relating large piezoelectricity to triple points are valid only for systems exhibiting inter-ferroelectric transformation, which is not the case in our non-conventional MPB system. At the same time, the softening and vanishing of the $A_1$(1TO) near the depolarization do confirm that the depolarization event is associated with lattice instability, although this does not lead to a ferroelectric-paraelectric transformation on the global scale as in normal ferroelectrics.

***Comparison of depolarization mechanisms in other systems***

The fact that the system shows a weak dielectric anomaly at the depolarization temperature during the heating cycle of the poled specimen suggests that the depolarization, inspite of not being associated with a thermodynamic phase transition as in normal ferroelectrics, should be triggered by some kind structural instability. A similar weak dielectric anomaly at ~ 70 $^{o}$C, i.e. well below the dielectric maximum temperature (150 $^{o}$C) has been reported during heating of a poled sample of the MPB composition of $PbTiO_3$-$Pb(Mg_{1/3}Nb_{2/3})O_3$ [70] and $(Ba,Ca)(Ti,Zr)O_3$ [71]. The anomalies in $PbTiO_3$-$Pb(Mg_{1/3}Nb_{2/3})O_3$ and $(Ba,Ca)(Ti,Zr)O_3$ have been attributed to temperature driven rhombohedral to tetragonal and orthorhombic-tetragonal inter-ferroelectric structural transformation, respectively. The $A_1$(1TO) mode survives above the depolarization temperature in $PbTiO_3$-$Pb(Mg_{1/3}Nb_{2/3})O_3$ since the system is still in ferroelectric phase of another symmetry (tetragonal) above the depolarization temperature. From the above, we can argue that although the temperature dependence of the dielectric behaviour of poled specimens of conventional MPB systems (exhibiting coexistence of tetragonal and rhombohedral/monoclinic symmetries) and the non-conventional MPB such as ours, may mimic similar behaviour, the mechanisms associated with the weak dielectric anomalies in the poled specimens in these systems are fundamentally different. In our case, (and perhaps in all non-conventional MPBs in this alloy category) there is no visible sign of the change in structural symmetry across the depolarization temperature. The vanishing of the $A_1$(1TO) mode at the depolarization temperature is analogous to ferroelectric-paraelectric thermodynamic transformation. The dramatic loss of piezoelectricity on annealing the poled specimen system at the depolarization temperature seems to support this argument. The piezoelectricity is not likely to decrease to such an extent if the depolarization is associated with an inter-ferroelectric instability as in $PbTiO_3$-$Pb(Mg_{1/3}Nb_{2/3})O_3$ and $BaTiO_3$-based systems. In the case of NBT, Rao et al [46] have demonstrated that the structural instability associated with depolarization is related to the onset of in-phase octahedral tilt – a structural distortion which is incompatible with long range ferroelectric order and therefore breaks the long-range coherence in polarization induced by the poling field.

## V. CONCLUSIONS

We have examined the mechanism associated with the large piezoelectric response in the low-Pb piezoelectric alloy system $PbTiO_3$-$Bi(Ni_{1/2}Zr_{1/2})O_3$ (PT-BNZ), the critical composition (x = 0.41) of which exhibit piezoelectric response comparable to the MPB compositions of PZT, both in the weak-field and large field regime. We demonstrate that, unlike the conventional MPB systems such as PZT, the two-phase state in our system are do not correspond to thermodynamically stable ferroelectric phases. Instead, the composition exhibiting large piezoelectric response in our system comprise of tetragonal regions of long and short-range spatial coherence. On the global length scale, this manifest as coexistence of tetragonal and cubic-like phases, a scenario mimicking MPB, but not so actually. Poling-field irreversibly suppresses the cubic-like phase, making the system appear as single phase tetragonal. We found a strong correlation between the measured tetragonality and the volume fraction of the phase in the unpoled state of the specimen. On poling, i.e., after suppression of the cubic-like phase, the tetragonality increases considerably for the same composition. This suggest that the measured lattice parameters from the XRD studies correspond to tetragonal regions of different spatial coherence. On heating the system above room temperature, a lattice instability sets in well before the dielectric maximum temperature, and disrupts the poling-field induced enhancement in the structural coherence. In the absence of a genuine inter-ferroelectric instability, it is not possible to invoke field induced polarization-rotation and inter-ferroelectric transformation to explain the large piezoelectricity in our case. Our study proves that large piezoelectricity is realizable even in non-MPB alloy systems by enhancing domain wall mobility. In this context, we argue that the alloy category represented by the general formula $PbTiO_3$-$Bi(M'M'')O_3$ are promising systems for design of new generation of non-MPB based high performance piezoelectric materials. These alloys have the added advantage that they have significantly reduced Pb content and can serve as possible alternatives to conventional all-PB MPB alloys.

## ACKNOWLEDGMENT

R. Pandey gratefully acknowledges the Science and Engineering Research Board (SERB) of the Department of Science and Technology, Government of India, for financial support as National Postdoctoral Fellowship (PDF/2015/000169). R. Ranjan gratefully acknowledges the SERB of the Ministry of Science and Technology, Government of India, for financial support (Grant No. EMR/2016/001457).

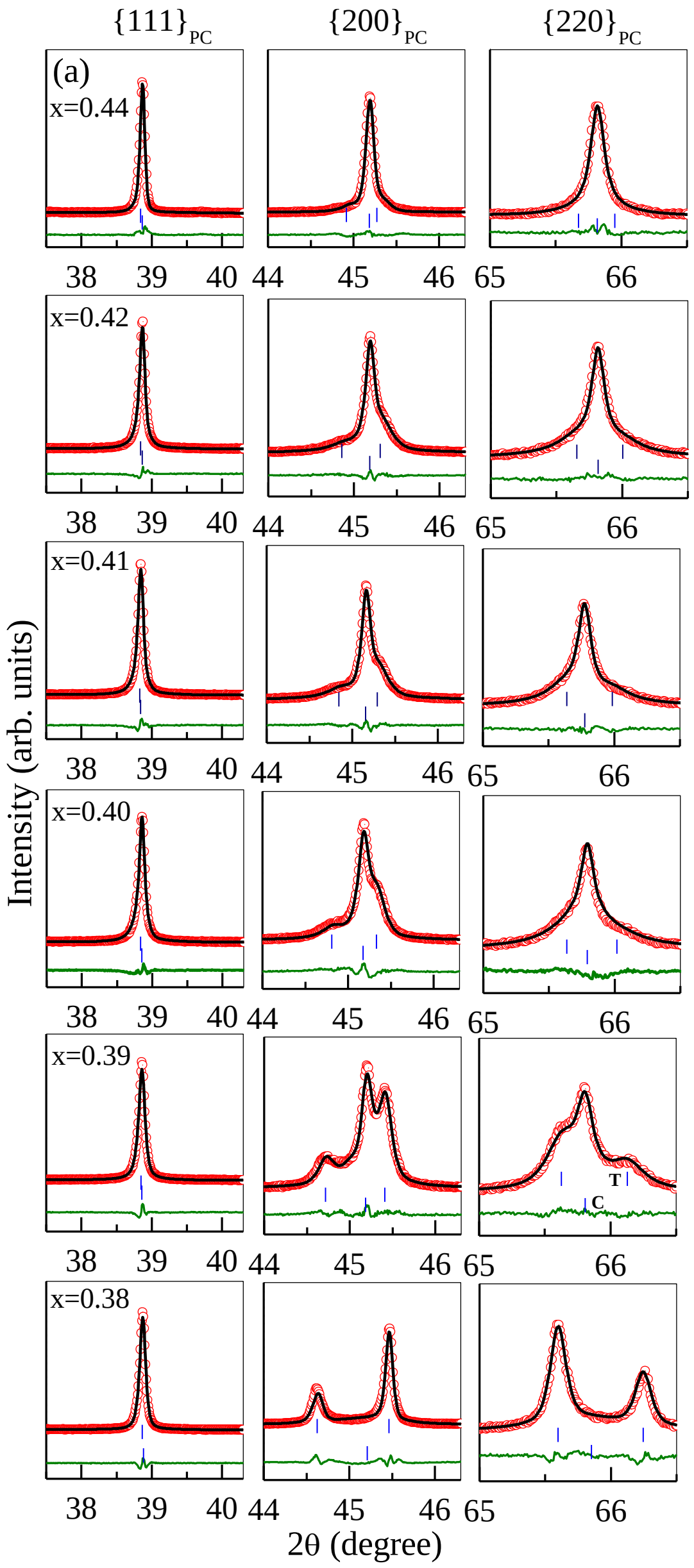

.

**Figure 1:** Observed (circles), calculated (continuous line) and difference (continuous bottom line) Rietveld-fit of XRD profile for the compositions with x = 0.38 - 0.44 taking coexisting cubic (*Pm*-3*m*) and tetragonal (*P*4*mm*) structures for (1-x)PT-xBNZ. The vertical tick-marks show the position of Bragg-peaks

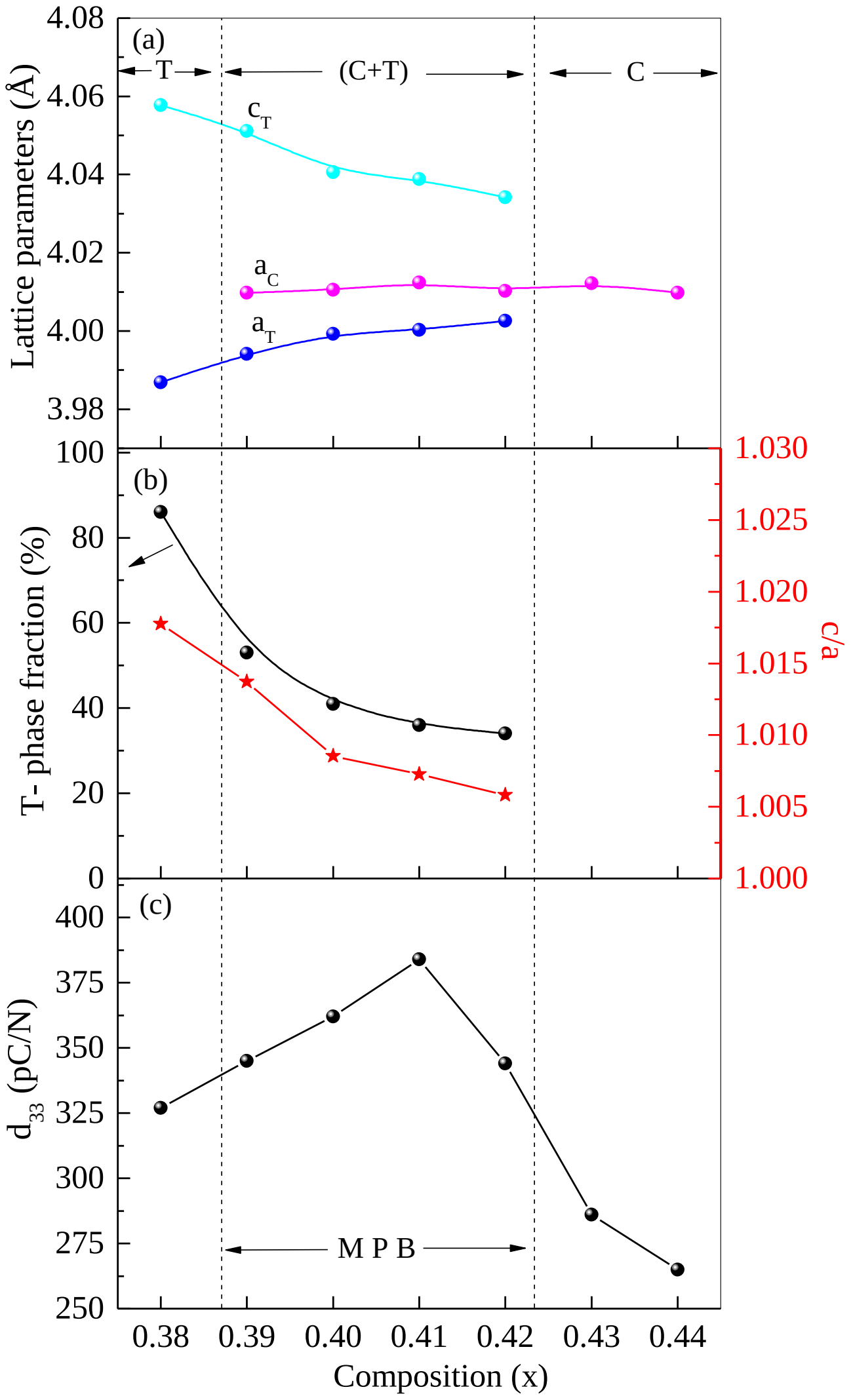

**Figure 2: (a)** Composition dependence of the cubic-like lattice parameter ($a_C$), and tetragonal lattice parameters ($c_T$ and $a_T$) of (1-x)PT-xBNZ. **(b)** Composition dependence of tetragonal phase fraction and the tetragonality. **(c)** Composition dependent longitudinal piezoelectric coefficient ($d_{33}$) for (1-x)PT-xBNZ. The vertical dotted lines denote the MPB region.

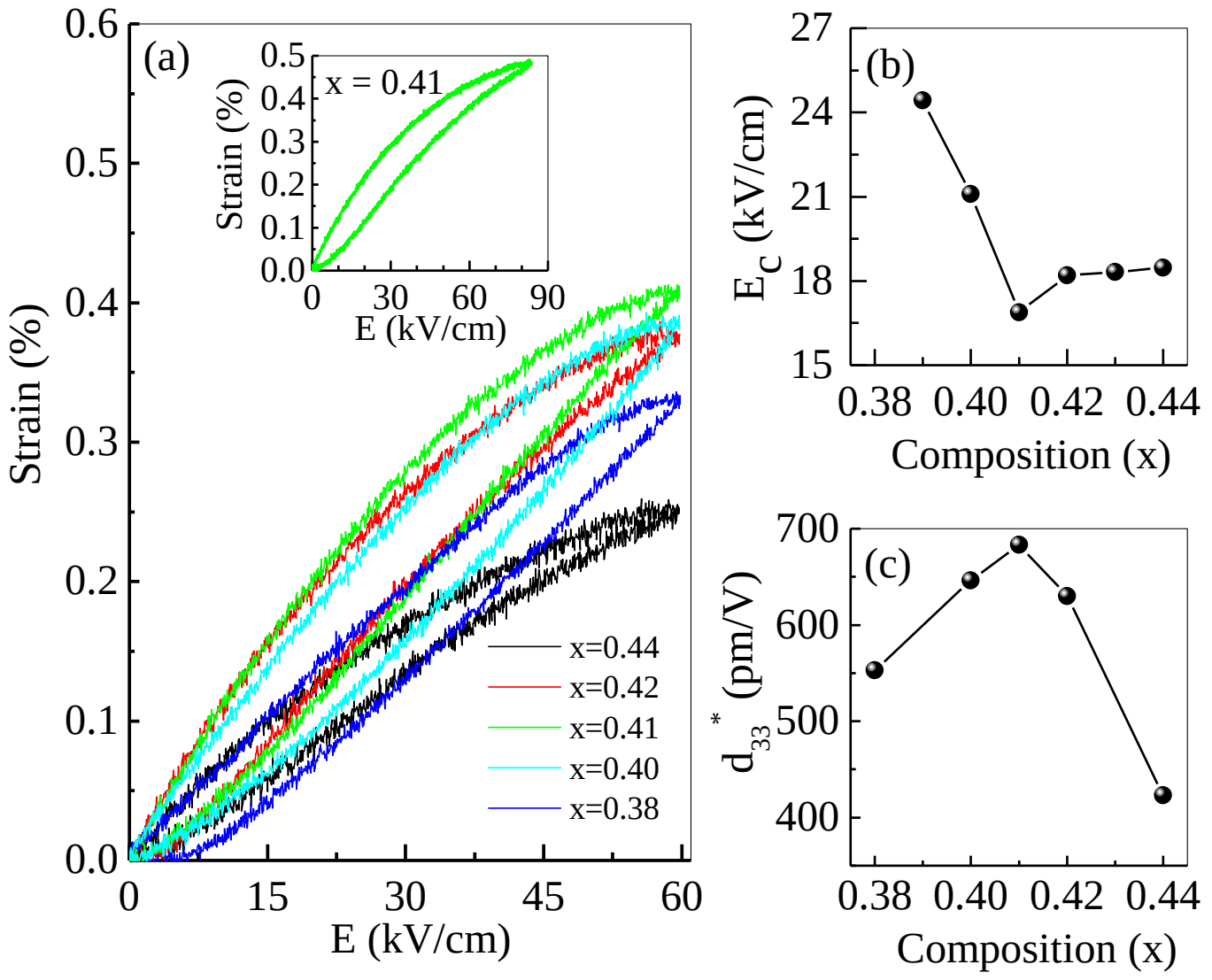

**Figure 3: (a)** Electric field (E) dependent unipolar strain (%) for (1-x)PT-xBNZ with x = 0.38 - 0.44. Inset shows the strain curve at applied field of 85 kV/cm for x=0.41 **(b)** Composition dependent coercive field, and **(c)** large signal piezoelectric coefficient ($d_{33}^*$).

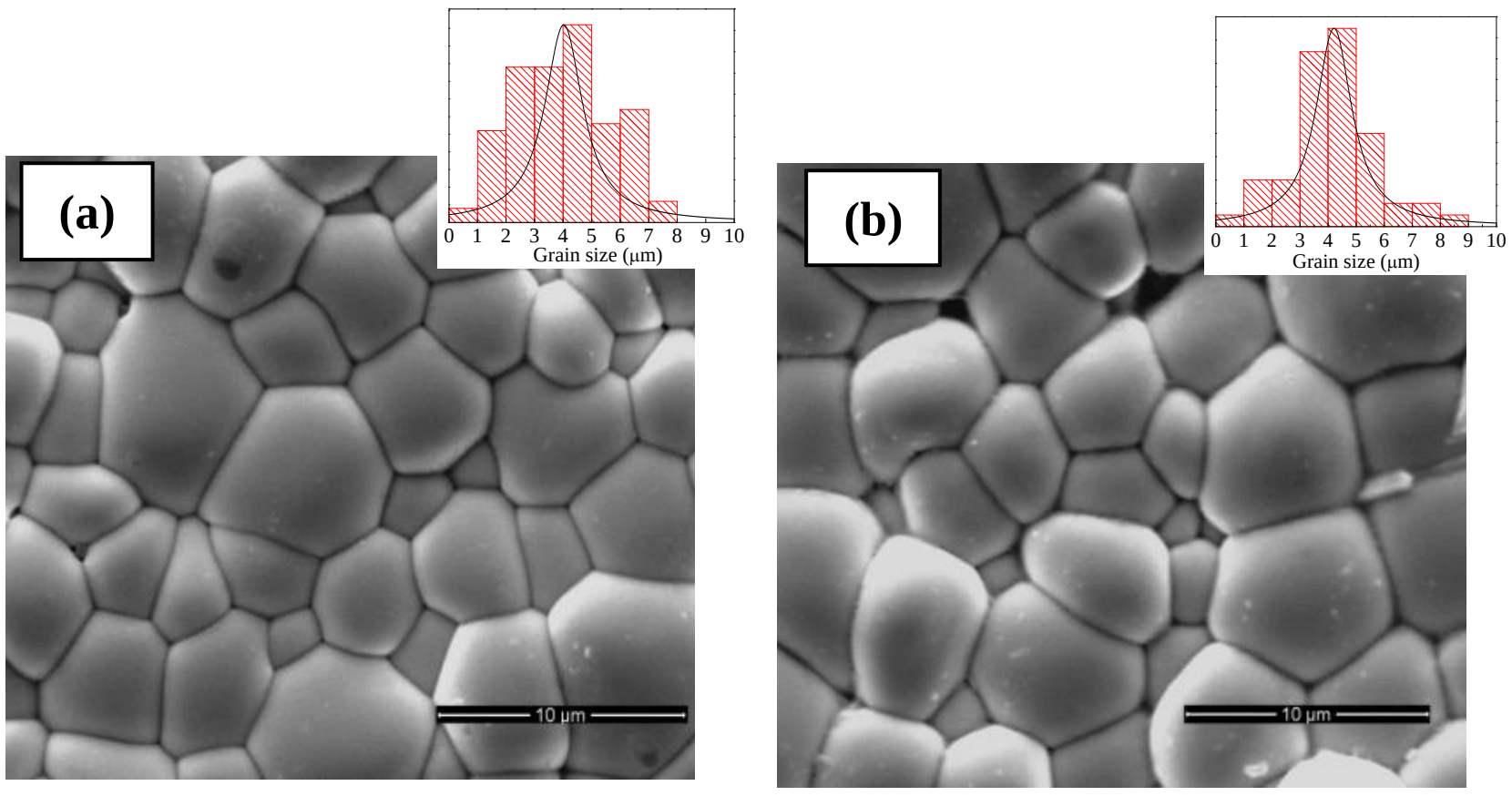

**Figure 4:** SEM images of (1-x)PT-xBNZ for **(a)** x = 0.41 and **(b)** x = 0.38. Grain size distribution is shown by the histogram on the SEM image.

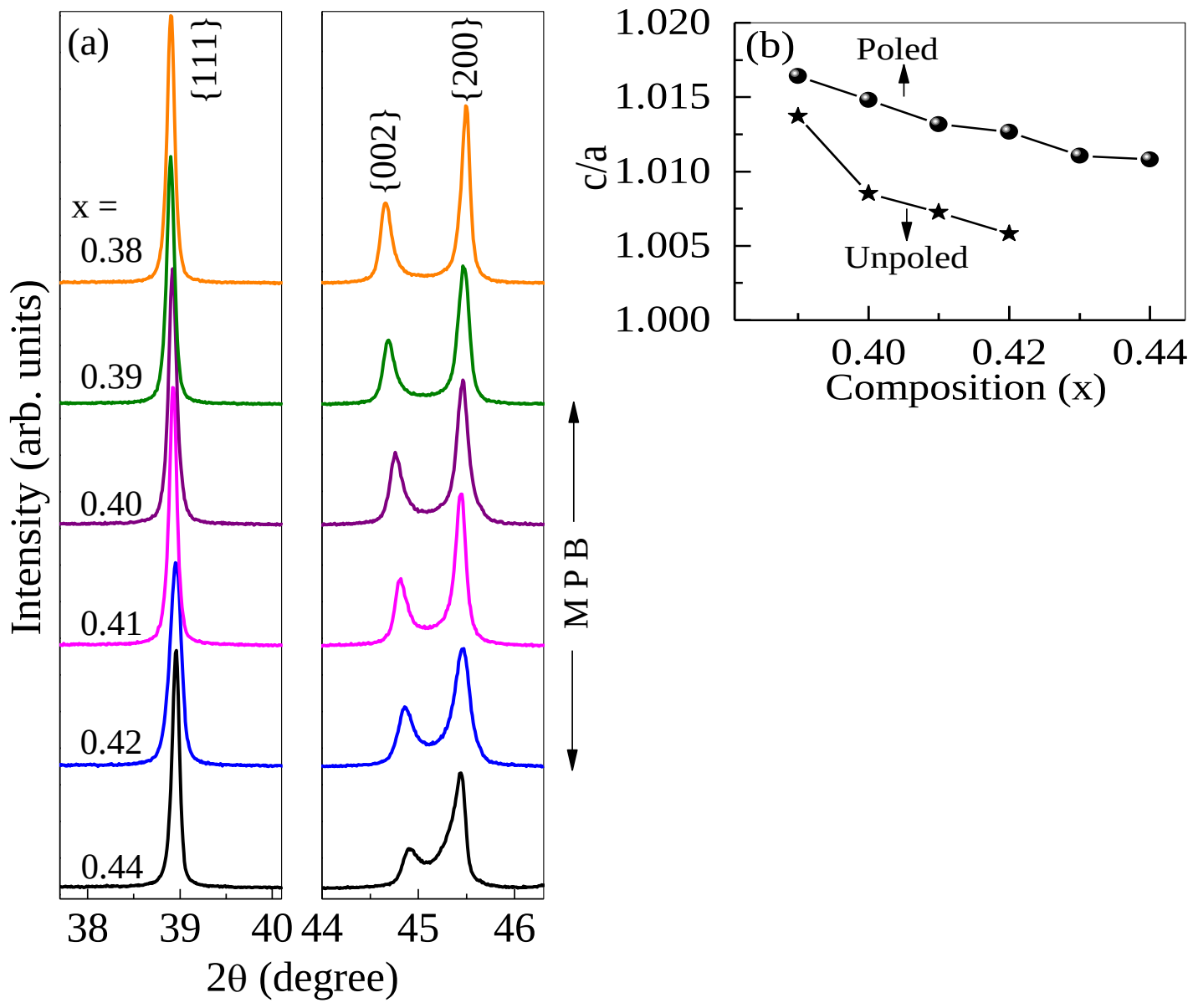

**Figure 5: (a)** {111}, and {002}/{200} powder XRD profiles of (1-x)PT-xBNZ for x = 0.38 - 0.44 poled at 30 kV/cm. Peaks are marked with tetragonal *P*4*mm* space group. **(b)** Composition dependent tetragonality of poled and unpoled (1-x)PT-xBNZ.

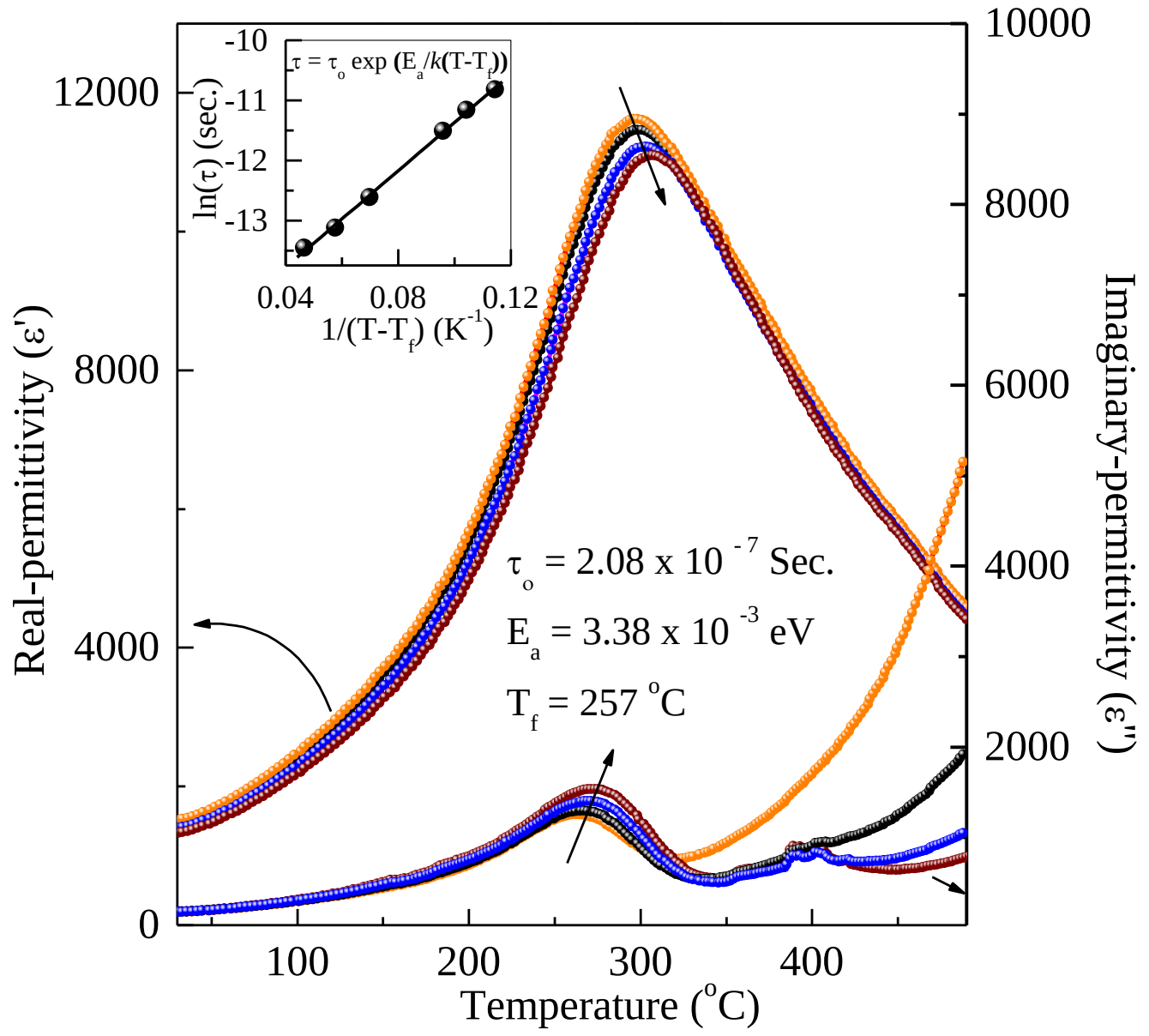

**Figure 6:** High temperature real part of permittivity (ε′) and imaginary part of permittivity (ε″) for PT-BNZ measured at 80 kHz, 100 kHz, 200 kHz and 400 kHz with x = 0.41. Inset shows the linear fit of the relaxation time (τ) for the Vogel-Fulcher freezing.

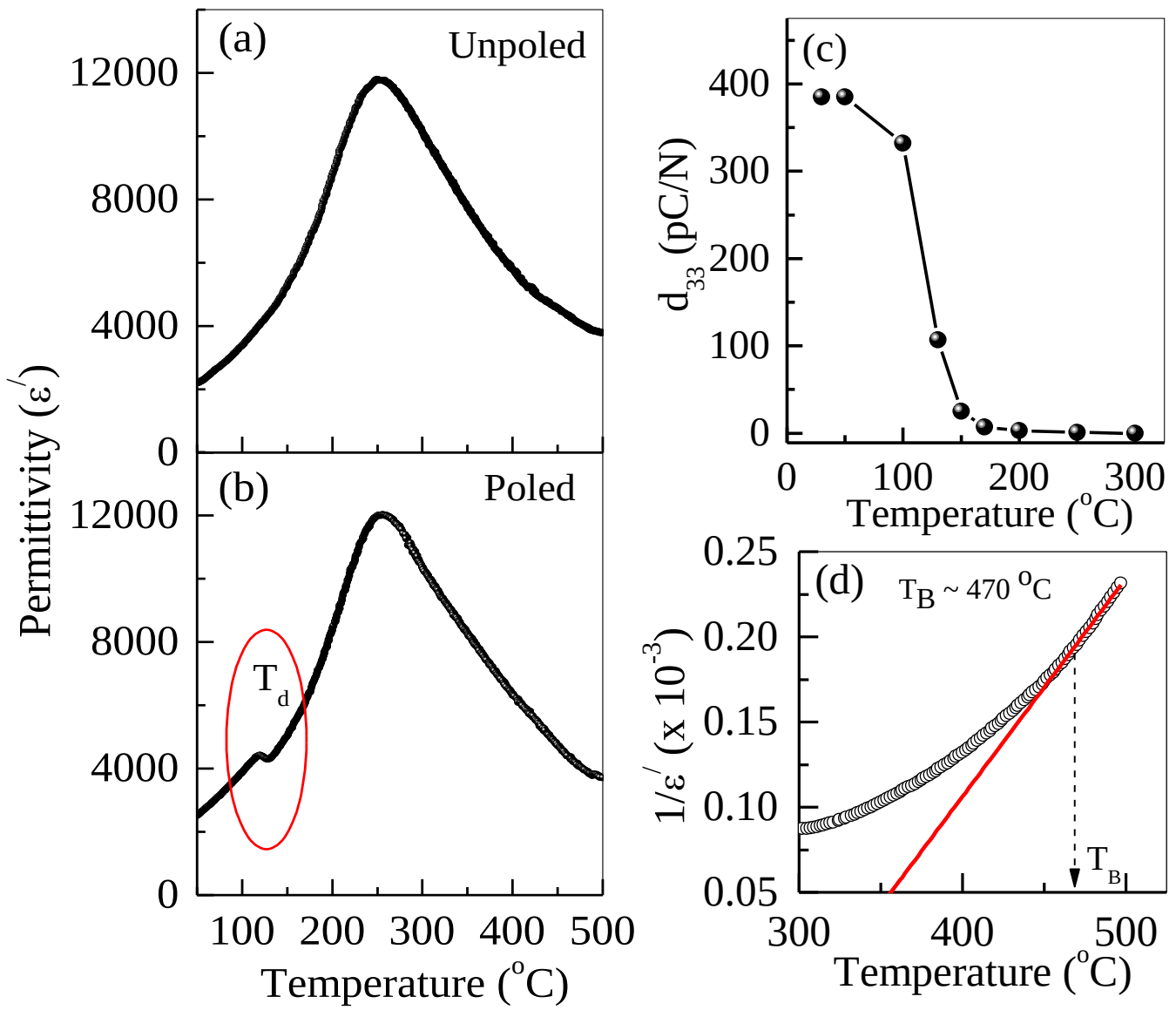

**Figure 7:** Temperature dependent real part of permittivity **(a)** unpoled **(b)** poled at 30 kV/cm; for PT-BNZ with x = 0.41 measured at 50 kHz. $T_d$ corresponds to depolarization temperature. **(c)** Variation of $d_{33}$ with annealing temperature. **(d)** $1/\varepsilon^{/}$ vs T plot and Curie-Weiss fit of the real part of the permittivity for PT-BNZ, x = 0.41. $T_B$ corresponds to Burn Temperature**.**

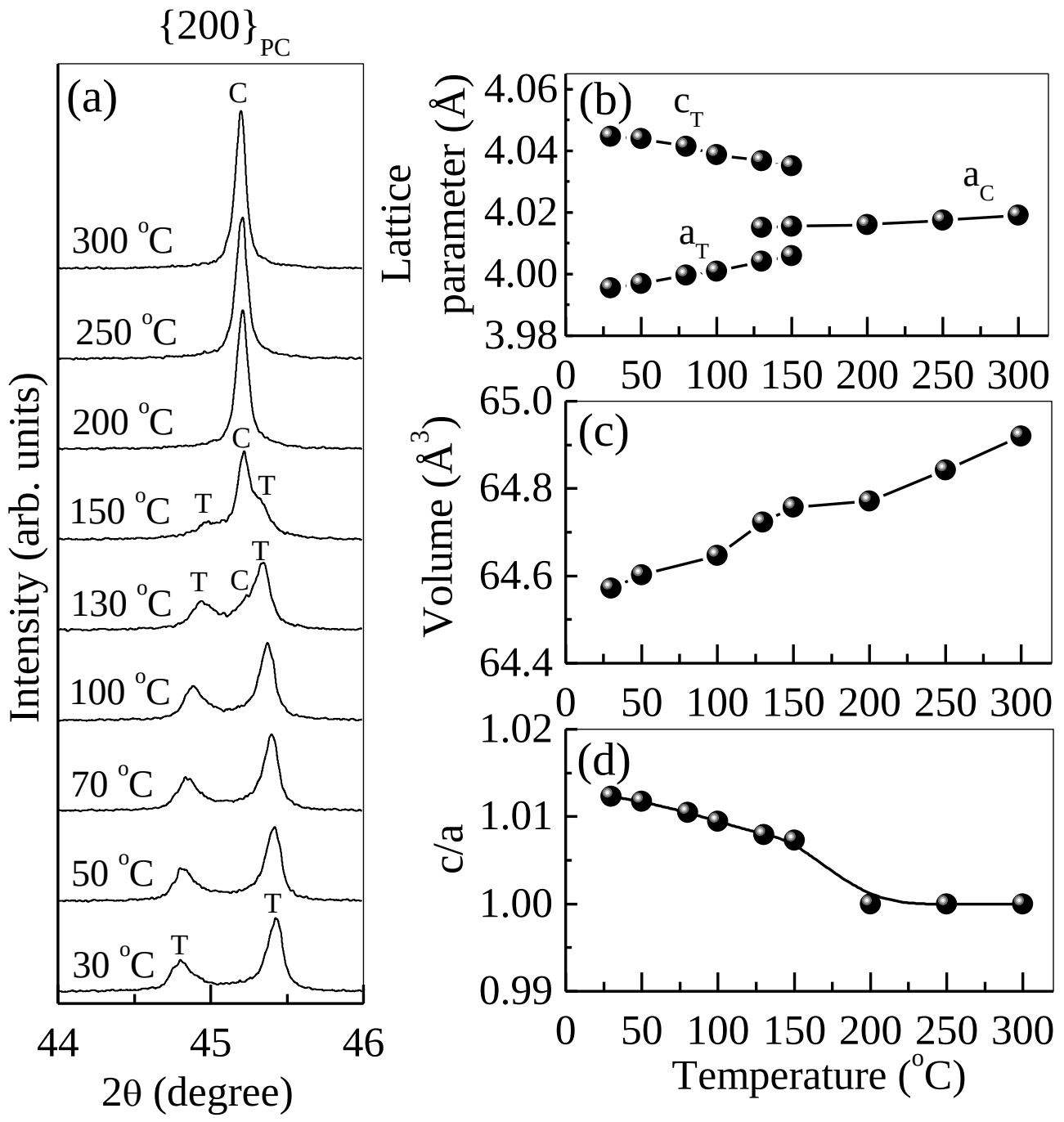

**Figure 8:** **(a)** Pseudo-cubic {200} XRD profile **(b)** variation of lattice parameter **(c)** unit cell volume, and **(d)** tetragonality (c/a); for poled PT-BNZ with x = 0.41 in the temperature range of 30 °C - 300 °C.

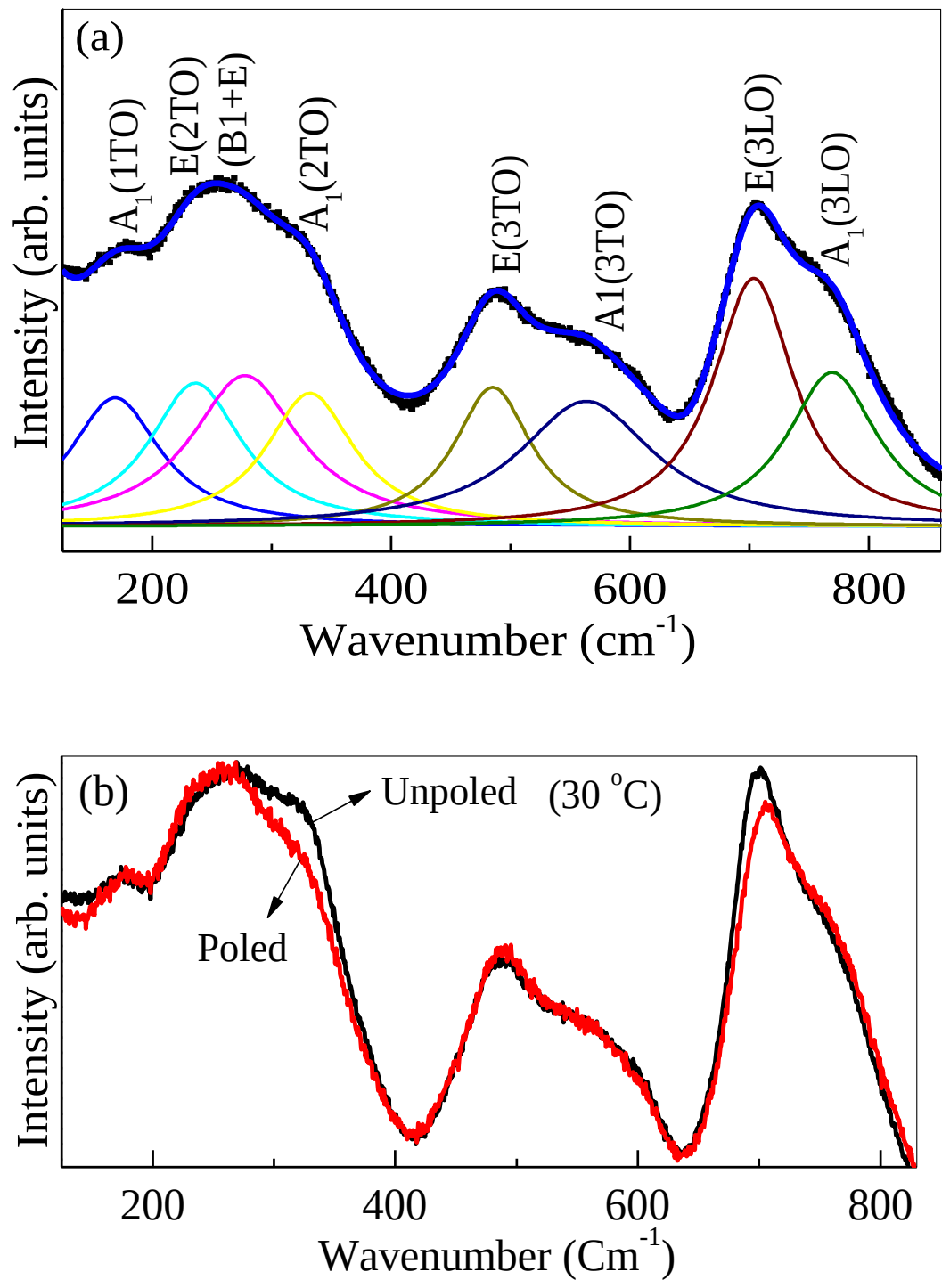

**Figure 9: (a)** Lorentzian curve fitting of Raman peaks using tetragonal structure (*P*4*mm*) for PT-BNZ (x = 0.41) poled at 30 kV/cm. **(b)** Comparison of the Raman spectrum at room temperature (30 °C) for the poled and unpoled samples.

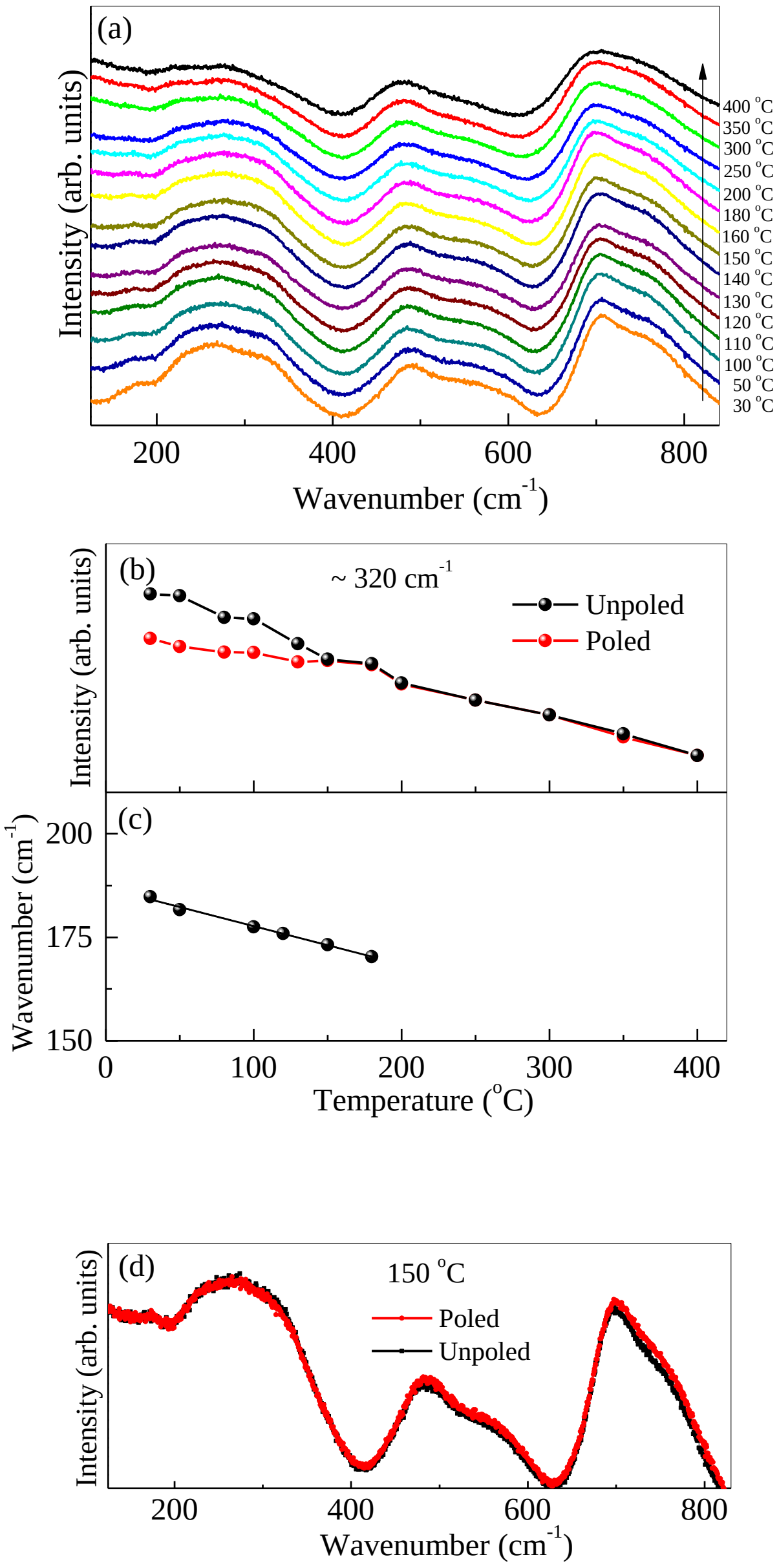

**Figure 10: (a)** High temperature (30 $^{o}$C - 400 $^{o}$C) Raman spectrum of PT-BNZ (x = 0.41) poled at 30 kV/cm. **(b)** Intensity of Raman active modes ~ 320 $cm^{-1}$ for poled and unpoled PT-BNZ (x = 0.41) during heating and cooling cycles. **(c)** Temperature dependence of $A_1$(1TO) Raman mode ~ 180 $cm^{-1}$. **(d)** Comparison of the Raman spectrum at 150 $^{o}$C of poled and unpoled samples.

**Table.1**. Refined structural parameters and agreement factors for PT-BNZ with x = 0.41 using tetragonal *P*4*mm* and cubic *Pm*-3*m* space groups.

| Composition | Ions | *P*4*mm* | *Pm*-3*m* |
|---|---|---|---|
| **x=0.41** | $Bi^{3+}/Pb^{2+}$ | x=y=z=0, $U_{11}=U_{22}$=0.043(8)Å$^2$, $U_{33}$=0.014(1)Å$^2$ | $B_{iso.}$=3.0(0)Å$^2$ |
| | $Ni^{2+}/Zr^{4+}/Ti^{4+}$ | x=y=0.5, z=0.5562(7), $B_{iso}$=0.1(5)Å$^2$ | $B_{iso.}$=0.2(1)Å$^2$ |
| (Unpoled) | $O^{2-}{}_{I}$ | x=y=0.5, z=0.103(2), $B_{iso}$=1.0(0)Å$^2$ | $B_{iso.}$=0.6(1)Å$^2$ |
| | $O^{2-}{}_{II}$ | x=0.5,y=0.0,z=0.637(2),$B_{iso}$=0.8(1)Å$^2$ | |
| | | $a_T$=4.0003(1)Å, $c_T$=4.0388(2)Å, | $a_C$=4.0122(3)Å |
| | | $\chi^2$=1.54, $R_{wp}$=8.16 | |
| | $Bi^{3+}/Pb^{2+}$ | x=y=z=0, $U_{11}=U_{22}$=0.041(1)Å$^2$, $U_{33}$=0.028(1)Å$^2$ | |
| (Poled) | $Ni^{2+}/Zr^{4+}/Ti^{4+}$ | x=y=0.5, z=0.5534(6), $B_{iso}$=0.4(1)Å$^2$ | |
| | $O^{2-}{}_{I}$ | x=y=0.5, z=0.102(2), $B_{iso}$=1.0(0)Å$^2$ | |
| | $O^{2-}{}_{II}$ | x=0.5,y=0.0,z=0.629(2),$B_{iso}$=0.8(1)Å$^2$ | |
| | | $a_T$=3.9982(6)Å, $c_T$=4.04581(1)Å, | |
| | | $\chi^2$=1.96, $R_{wp}$=8.94 | |